# Layer-Dependent Topological Phase in a Two-Dimensional Quasicrystal and Approximant

Jeffrey D. Cain[1,2,3,†], Amin Azizi[1,3,†], Matthias Conrad[4], Sinéad M. Griffin[2,5], and Alex Zettl[1,2,3*]

[1]*Department of Physics, University of California at Berkeley, Berkeley, CA 94720, USA*

[2]*Materials Sciences Division, Lawrence Berkeley National Laboratory, Berkeley, CA 94720, USA*

[3]*Kavli Energy NanoSciences Institute at the University of California at Berkeley and the Lawrence Berkeley National Laboratory, Berkeley, CA 94720, USA*

[4]*Fachbereich Chemie, Philipps-Universität Marburg, Hans-Meerwein-Straße 4, 35032 Marburg, Germany*

[5]*The Molecular Foundry, Lawrence Berkeley National Laboratory, Berkeley, CA 94720, USA*

*Corresponding author: azettl@berkeley.edu
†These authors contributed equally

**One Sentence Summary:** Isolation of a two-dimensional chalcogenide quasicrystal and approximant that possesses a layer tunable, topologically nontrivial band structure.

**Electronic and topological properties of materials are derived from the interplay between crystalline symmetry and dimensionality. Simultaneously introducing 'forbidden' symmetries *via* quasiperiodic ordering with low-dimensionality into a material system promises the emergence of new physical phenomena. Here, we isolate a two-dimensional chalcogenide quasicrystal and approximant, and investigate associated electronic and topological properties. Ultra-thin two-dimensional layers of the materials with a composition close to $Ta_{1.6}Te$, derived from a layered transition metal dichalcogenide, are isolated with standard exfoliation techniques and investigated with electron diffraction and atomic-resolution scanning transmission electron microscopy. Density functional theory calculations and symmetry analysis of the large unit-cell crystalline approximant of the quasicrystal $Ta_{21}Te_{13}$ reveal the presence of symmetry protected nodal crossings in the quasicrystalline and approximate phases whose presence is tunable by layer number. Our study provides a platform for the exploration of physics in quasicrystalline low-dimensional materials and the interconnected nature of topology, dimensionality and symmetry in electronic systems.**

In recent decades, two discoveries have caused dramatic shifts in the descriptions of order in solid state systems: the prediction and subsequent observation of topological order, and the discovery of Bragg diffraction in aperiodic quasicrystals. Topological phases are defined by geometric phases of the underlying wave function which result in unusual properties, such as boundary states that are robust to disorder. Interest in topological matter has surged since the discoveries of an ever-expanding collection of materials with symmetry-protected features.

Experimentally-realized topological materials include topological[1–3] and Chern insulators and Dirac and Weyl semimetals[4–8].

A similar paradigm shift was brought about by the discovery, nearly four decades ago, of aperiodic order and so-called 'forbidden' symmetries in quasicrystalline materials[9]. Quasicrystals occur when a structure has long-range order which is not compatible with long-range translational symmetry. This incompatibility arises from the presence of specific symmetries (*e.g.* 5-, 8, 10-, 12-fold) which are not compatible with translational symmetry (unlike, for example, 4- or 6-fold symmetries). Examples of aperiodic tilings include pentagonal, dodecagonal, and Penrose tilings[10]. First demonstrated in quenched aluminum alloys[9], quasicrystalline order has been observed in other metallic systems[11], self-assembled colloidal crystals[12], ultracold atoms[13], and CVD grown bilayer graphene[14–16].

Recent work has reexamined what classes of systems can host topological phenomena, and the catalogue has expanded to include amorphous materials[17,18] and light elements[19]. Additionally, several theoretical works have examined the influence of aperiodicity and quasicrystallinity on topological phases including pentagonal[20] and octagonal[21] quasicrystals and so-called 'higher-order' topological insulators[22]. Further, topological polarization not possible in periodic crystals has been theoretically proposed to exist in quasicrystals. Artificial quasiperiodic systems have been created experimentally in photonic crystals for which a photonic waveguide is used to construct a 1D Hamiltonian with 2D topological edge modes[23]. Two-dimensional quasicrystals have specifically been predicted to host novel electronic states including Hofstadter's Butterfly[24] and the four-dimensional quantum Hall effect[25], but to date no specific quasicrystalline solid state system has been proposed or discovered that can host nontrivial electronic topological states. There are examples of low-dimensional quasicrystalline solids in the literature, for example CVD grown 30° bilayer graphene[26] and 2D $BaTiO_3$ epitaxially grown on Pt[27]. However, quasicrystallinity has not been observed in exfoliated materials, limiting the exploration of quasicrystalline order in 2D materials and van der Waals heterostructures.

In the binary tantalum-tellurium system, a layered-type quasicrystalline phase exhibiting a twelvefold symmetric diffraction pattern was discovered.[28] The composition of this dodecagonal (dd) phase is close to $Ta_{1.6}Te$. In intimate proximity to dd-$Ta_{1.6}Te$ exist several crystalline phases, so-called approximants, that are compositionally and structurally closely related to the quasicrystalline compound. The structures of two approximants, $Ta_{97}Te_{60}$[29] and $Ta_{181}Te_{112}$[30], were determined by means of single crystal X-ray diffraction, whereas $Ta_{21}Te_{13}$[30] was identified by means of electron diffraction and high-resolution electron microscopy. Based upon the knowledge of the structures of $Ta_{97}Te_{60}$ and $Ta_{181}Te_{112}$ the building principle underlying the approximant structures was deciphered. This building principle allows for the derivation of structure models of new approximants as was shown for $Ta_{21}Te_{13}$. Furthermore, the algorithm can be used for generating structure models for the quasicrystalline phase.

Here, we expand the catalogue of exfoliated two-dimensional materials to include quasicrystals and approximants; specifically, the van der Waals material dd-$Ta_{1.6}Te$[28] and $Ta_{181}Te_{112}$ approximant[30]. We isolate few-layers of the materials and investigate their structure using electron diffraction and atomic-resolution scanning transmission electron microscopy. Using first-principles calculations and symmetry analysis performed on the large unit-cell crystalline approximant, $Ta_{21}Te_{13}$,[30] we calculate the material's electronic structure and find that it can host symmetry-protected nodal states, whose presence is dependent upon layer number.

The material is prepared by reducing the transition metal dichalcogenide $TaTe_2$[31] at high temperature with Ta, as previously reported by one of the authors in Conrad *et. al.*[28] (See

Supplementary Information for synthesis details). The resulting material can be a mixture of the dd-phase and approximant ($Ta_{181}Te_{112}$). The processing induces a step-wise transformation from the monoclinic $TaTe_2$ structure (Fig. 1a) into the dd quasicrystalline form dd-$Ta_{1.6}Te$ and $Ta_{181}Te_{112}$. Common to the dd phase and approximants are twelvefold Ta-Te clusters, shown in Fig. 1b. Both compounds exhibit anisotropic structures, with out-of-plane (periodic) Te terminated layers, separated by a van der Waals gap, analogous to the transition metal dichalcogenides (*e.g.* $MoS_2$, $WSe_2$). For both the dd phase and approximant, the layers are approximately 1 nm thick (Fig. 1c). The resulting crystals are 100's of microns in diameter, and exhibit platelet-like morphology with a metallic gold luster and an obvious lamellar, lubricant-like structure. The resulting materials are unstable in ambient atmosphere and are stored and handled in an inert atmosphere (See Supplementary Information for details). Electron microscopy sample preparation details, along with optical images of the transferred material, are presented in the SI (Fig. S1). Electron diffraction of the dd-$Ta_{1.6}Te$ is shown in Fig. 1d. The diffraction pattern exhibits twelvefold symmetric spots (red dodecagon), separated by 30°. The pattern also shows the self-similar, multiscale, twelvefold symmetric structure (green and yellow dodecagons) characteristic of quasicrystalline materials. An electron diffraction pattern from the approximant phase, $Ta_{181}Te_{112}$, is shown in Fig 1e; the periodic superstructure is easily seen in the inset, indicating the presence of periodicity and translational symmetry in the lattice. As noted above, the samples of the materials are lamellar, and easily cleaved with the "scotch tape" method. This is demonstrated by the optical image in Fig. 1f which shows that the materials can be exfoliated into two-dimensional building blocks. Two-dimensional samples of $Ta_{1.6}Te$ are identified by optical contrast and ultra-thin samples are readily produced using standard exfoliation techniques (Fig. S2). Further, Fig. S3 shows a low-magnification annular dark-field scanning transmission electron microscope (ADF-STEM) image, which shows the discrete layers of the exfoliated material.

The structure of $Ta_{181}Te_{112}$ is next investigated using electron microscopy and atomic-resolution STEM imaging. First, the composition is mapped in Fig. 2a using energy dispersive X-ray spectroscopy (EDS), along with a reference image, which shows homogenous distribution of Ta and Te (see the EDS spectrum in Fig. S4). Fig. 2b shows a low-magnification ADF-STEM image of the material, in which twelvefold Ta-Te clusters are evident. From the image, it is not immediately evident which phase is present; therefore, the inset shows a Fourier transform of the image in which the periodic superstructure (square pattern of dots) of the approximant can be clearly seen. Further, high-magnification images of a Ta-Te cluster (top: unfiltered, bottom: filtered) are shown in Fig. 2c. The structures of dd-Ta $_{1.6}$Te and its approximants can be characterized by square-triangle tilings on different length-scales. The basic or secondary tiling consists of tiles with an edge length of about 0.5 nm, which are shown together with their decoration with Ta- and Te-atoms in Fig. 3a. These tiles are then combined into the dodecahedral motifs, yielding squares and triangles with an edge length of about 2 nm presented in Fig. 3b (schematic and ADF-STEM images). These tiles span the so-called tertiary tiling. If this algorithm is repeated ad infinitum we obtain a self-similar square-triangle tiling which was first described by Stampfli[31], and may be used as a model for the quasicrystalline phase. Note that the dodecahedral motifs can occur in two orientations which are related by a rotation about 30°. The same tiles with 2 nm edge length are used to tile both the dd and approximant structures, with local atomic structure shared in both phases. Figure 3c shows a partially tiled image of $Ta_{181}Te_{112}$ and rather than tiling in an aperiodic fashion of the dodecagonal structure a periodic tiling is achieved. The unit cell of the approximant is highlighted in purple. In this image, all of the triangle tiles have the same local atomic structure, and all of the square tiles have the same local atomic structure. Tiles with the same color have the same orientation within the image.

To investigate the electronic structure of the dd-Ta$_{1.6}$Te quasicrystal and its approximant, as well as their evolution upon going from three dimensions to two, we consider another approximant material, Ta$_{21}$Te$_{13}$, so that we may apply density functional theory methods to the system, which require Bloch periodicity. Full calculation details are given in the SI. We emphasize while the in-plane structural tiling in the ideal quasicrystal is aperiodic, it is periodic in the out-of-plane direction. Approximants of quasicrystals are periodic crystals whose local structural motifs match those of the quasicrystal and that, within large unit cells, reproduce a portion of the quasiperiodic structure. Ta$_{21}$Te$_{13}$ as a platform for calculating the electronic properties of dd-Ta$_{1.6}$Te and Ta$_{181}$Te$_{112}$ is a reasonable choice for several reasons: locally, it has 12-fold symmetric motifs common to both (Shown in Fig. 4a and b), the local bonding is consistent, and finally, its unit cell is large enough to capture the relevant physics of the quasicrystalline phase, and the super-large approximant Ta$_{181}$Te$_{112}$, yet small enough to be feasible for full *ab initio* treatment. We note that the use of approximants for the theoretical investigation of quasicrystals has been explored, for example in twisted graphene bilayers, and their applicability proven by direct comparison with full quasicrystal lattices[33,34].

We first classify the topological properties of three distinct structures comprising Ta$_{21}$Te$_{13}$ units to deduce the layer dependence of the electronic structure. We consider a bulk structure, a monolayer with a 10Å vacuum, and a bilayer with a 10Å vacuum. In all cases, the systems adopt *P6mm* space/layer group symmetry. Topological classification is carried out both with and without spin-orbit coupling (SOC) with the results summarized in Table I; the Brillouin zone of dd-Ta$_{1.6}$Te is provided for reference in Fig 4c.

**TABLE I. Summary of symmetry-protected topological properties of the $Ta_{21}Te_{13}$ for bulk, bilayer, and monolayer phases**. The topological phases included are high-symmetry-point semimetals (HSPSM) and high-symmetry-line semimetals (HSLSM). For the former, the high-symmetry point coordinates, and the corresponding bands are given, while for the latter the high-symmetry lines where symmetry-protected crossings appear are listed

|  | Spin-orbit Coupling? | Topological Classification | Position in Brillouin Zone | Bands | Degenerate Irreps |
|---|---|---|---|---|---|
| **Bulk** | No | HSPSM | A | 366-367 | $A_5$ |
|  |  |  | Γ | 366-367 | $Γ_5$ |
|  | Yes | HSLSM | Γ-A line | -- | -- |
| **Bilayer** | No | HSPSM | A | 732-733 | $A_5$ |
|  |  |  | Γ | 732-733 | $Γ_5$ |
|  | Yes | Trivial Insulator | -- | -- | -- |
| **Monolayer** | No | HSLSM | M-Γ, M-A, L-Γ, L-A, M-Γ | -- | -- |
|  | Yes | Trivial Insulator | -- | -- | -- |

We first discuss the case of bulk $Ta_{21}Te_{13}$. Without SOC we predict a high-symmetry point semimetal (HSPSM) where the valence and conduction bands meet at the Γ and A high-symmetry points. With SOC we find a high-symmetry line semimetal (HSLSM) which manifests when the compatibility relations between high-symmetry points in the Brillouin zone are not satisfied – when this occurs there is at least one topological node along the line connecting the high-symmetry points. Schematics of HSPSM and HSLSM band structures are shown in Fig. 4d. In our case, the node occurs on the Γ − A high-symmetry line, which is a two-fold degenerate crossing. To further investigate this, we calculate the electronic band structures of the *P6mm* bulk system with and without SOC as shown in Figs. 4e and g, respectively. Both phases are metallic with several band crossings throughout the Brillouin zone. However, many band degeneracies are lifted with SOC, as expected. We confirm the nodal points at Γ and A in the absence of SOC, and the nodal crossing along the Γ − A line with SOC (Fig. 4f).

The bilayer results without SOC are very similar to the bulk case. Our calculations give a HSPSM at Γ and A. However, upon inclusion of SOC, these crossings are gapped out and the system is reduced to a trivial insulator. We find similar behavior for the properties of the monolayer, where without SOC we find a HSLSM with nodal crossings throughout the Brillouin zone. These are gapped with the inclusion of SOC with the monolayer becoming a trivial insulator. Motivated by the observation that the bulk and layered compounds have different topological behavior upon the inclusion of SOC, we examine the topological properties of the bulk compound while changing the magnitude of the interlayer separation in the structure. We find that the bulk structure retains its topological character in varying the nearest inter-layer Te-Te separation from its experimental value (1.8 Å) up to 2.8 Å. Beyond this critical point, the bulk structure is topologically trivial. Finally, we calculate the electronic densities of states for the three structures (Fig. S5). All three have almost identical dispersion – transitioning from bulk to monolayer does not reduce band dispersion.

We next discuss the influence of SOC on the symmetry-protected crossings. SOC causes a spin splitting which can be clearly seen in Fig. 4g. In the bulk case, this splits the high-symmetry points to a crossing along the high-symmetry line. Moving to the monolayer and bilayer cases, SOC gaps out the nodes resulting in trivial insulators.

We find that the presence of topological nodal crossings strongly depends on interlayer separation. In our computer experiment we vary the interlayer separation in the bulk structure and find a critical value of 2.8 Å, above which the system is a trivial insulator. Physically, modifying the interlayer separation changes the interlayer coupling, and in our case it causes a topological phase transition to occur once the separation is large enough, *i.e.* when exfoliated into few/single layers. This suggests that modifying both the out-of-plane separation, or layer number, and the magnitude of SOC (by chemical substitution with lighter elements) can be used as tuning parameters for controlling the topological phase transitions in these layered materials.

We finally address whether our results for the $Ta_{21}Ta_{13}$ periodic approximant are also relevant to the dd phase and $Ta_{181}Te_{112}$. As previously discussed, the main structural changes in going from the approximant to the quasicrystal involves the connectivity of the local structural motifs. However, both phases contain the same local structural ordering and, importantly, the out-of-plane bonding is the same across structural variants. In addition to the arguments above, *i.e.* similar local structural motifs with appropriate length scales, we note that the symmetry-projected nodes occur in the out-of-plane $\Gamma-A$ direction. We confirm that modifications in the interlayer coupling can remove these nodes in the bulk phase. However, the aperiodicity in these systems manifests within the plane, and does not extend into the out-of-plane direction. Therefore, our topological nodal crossings should persist in the quasicrystalline and approximant phases.

In summary, we have isolated a two-dimensional chalcogenide dodecagonal quasicrystal and approximant, $Ta_{1.6}Te$. We investigated the structure of the approximant phase with atomic-resolution electron microscopy, and explored the topological properties of the materials as a function of layer number. Density functional theory calculations show the presence of a layer tunable topological band structures, not previously seen in solid state quasicrystalline systems. Our study lays the groundwork for the study of novel electronic states in low-dimensional quasiperiodic systems, the integration of quasicrystals into van der Waals heterostructures, and the interplay between topology, dimensionality, and symmetry.

**Acknowledgements**

Support was provided by the U.S. Department of Energy, Office of Science, Office of Basic Energy Sciences, Materials Sciences and Engineering Division under Contract No. DE-AC02-05-CH11231, under the sp$^2$-bonded Materials Program (KC2207), which provided for TEM characterization. Additional support was provided by the U.S. Department of Energy, Office of Science, Office of Basic Energy Sciences, Materials Sciences and Engineering Division under contract No. DE-AC02-05-CH11231, within the van der Waals Heterostructures Program (KCWF16), which provided for 2D material preparation and STEM characterization. The authors also gratefully acknowledge B. Harbrecht and his original work on the material, and W. Hornfeck for his valuable discussion and expertise regarding the material.

S.M.G. acknowledges the encouragement and guidance of Alexey Soluyanov throughout this work before his untimely passing. Computational resources were provided by the National Energy Research Scientific Computing Center and the Molecular Foundry, DOE Office of Science User Facilities supported by the Office of Science of the U.S. Department of Energy under Contract No. DE-AC02-05CH11231. The work performed at the Molecular Foundry was supported by the Office of Science, Office of Basic Energy Sciences, of the U.S. Department of Energy under the same contract.

**Author Contributions**

A.Z. and J.D.C. conceived of the experiments. M.C. synthesized the material. J.D.C. and A.A. prepared the samples, contributed to electron microscopy, and analyzed the images. S.M.G. performed first principles calculations.

**Author Information**

Corresponding author Alex Zettl

*Email: azettl@berkeley.edu

The authors declare no competing financial interest.

**Figures**

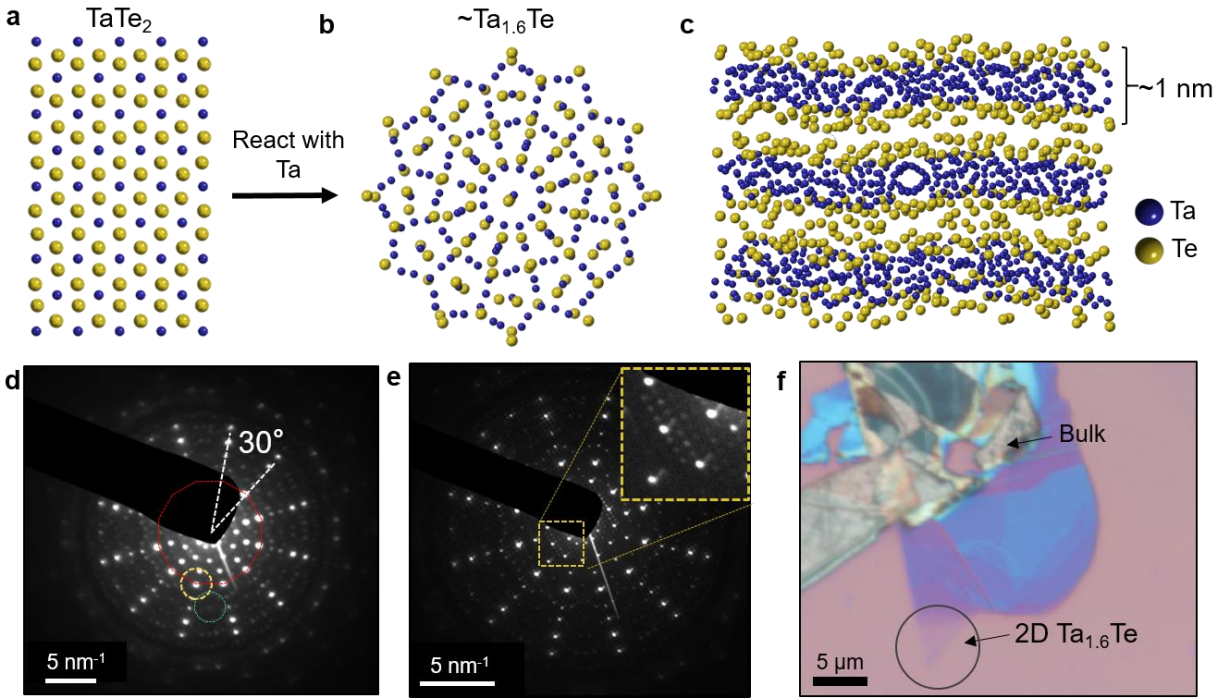

**Fig 1. Creation and structure of two-dimensional Ta$_{1.6}$Te. a**, TaTe$_2$ and **b**, Ta-Te cluster within dd-Ta$_{1.6}$Te and Ta$_{181}$Te$_{112}$ approximant. **c**, Side view of the layered structure showing the Te-Ta-Te slabs, 1 nm thick. **d**, Electron diffraction pattern of dd-Ta$_{1.6}$Te showing twelvefold symmetry (red) and self-similar motifs (green and yellow). **e**, Electron diffraction of the large unit-cell approximant Ta$_{181}$Te$_{112}$. Inset: Magnified area showing periodic superstructure. **f**, Optical microscope image of exfoliated Ta$_{1.6}$Te, two-dimensional section highlighted within black circle.

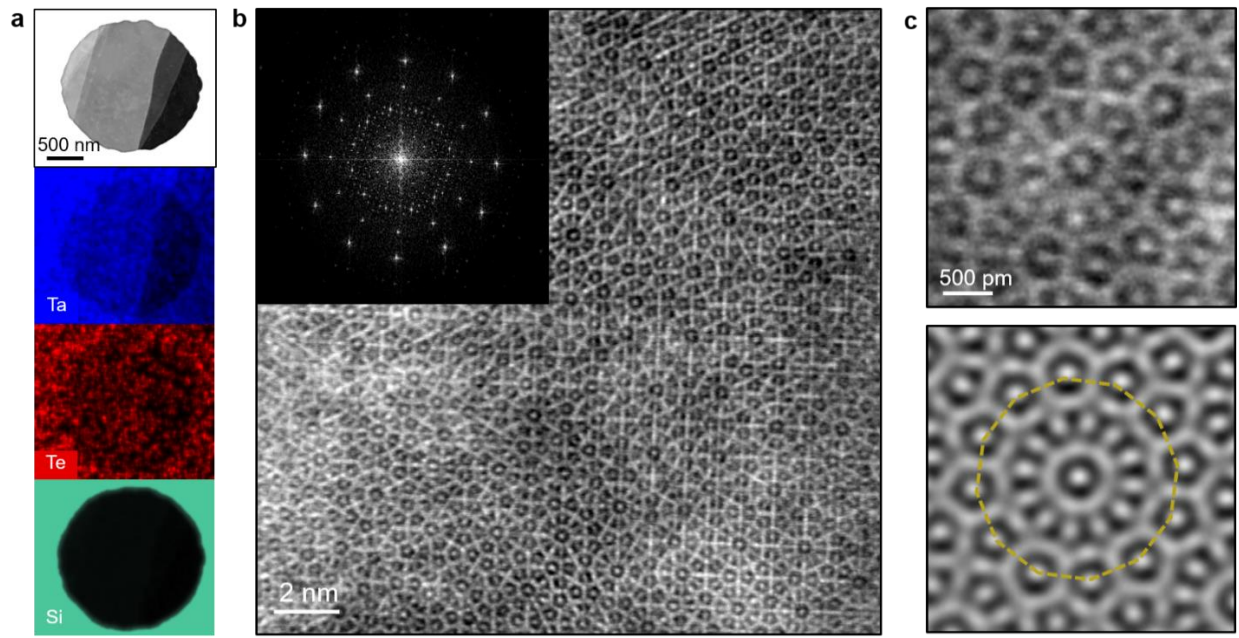

**Fig 2. High-Resolution imaging of Ta$_{181}$Te$_{112}$**. **a**, EDS maps and reference image showing uniform distribution of Ta and Te in the material (See Fig. S4 for the EDS spectrum). **b,** Low-magnification STEM image, inset: Fourier Transform of image. **c**, High-magnification image of Ta-Te cluster, top: unfiltered, bottom: filtered. The dashed circle highlight twelvefold symmetric nature of Ta-Te clusters.

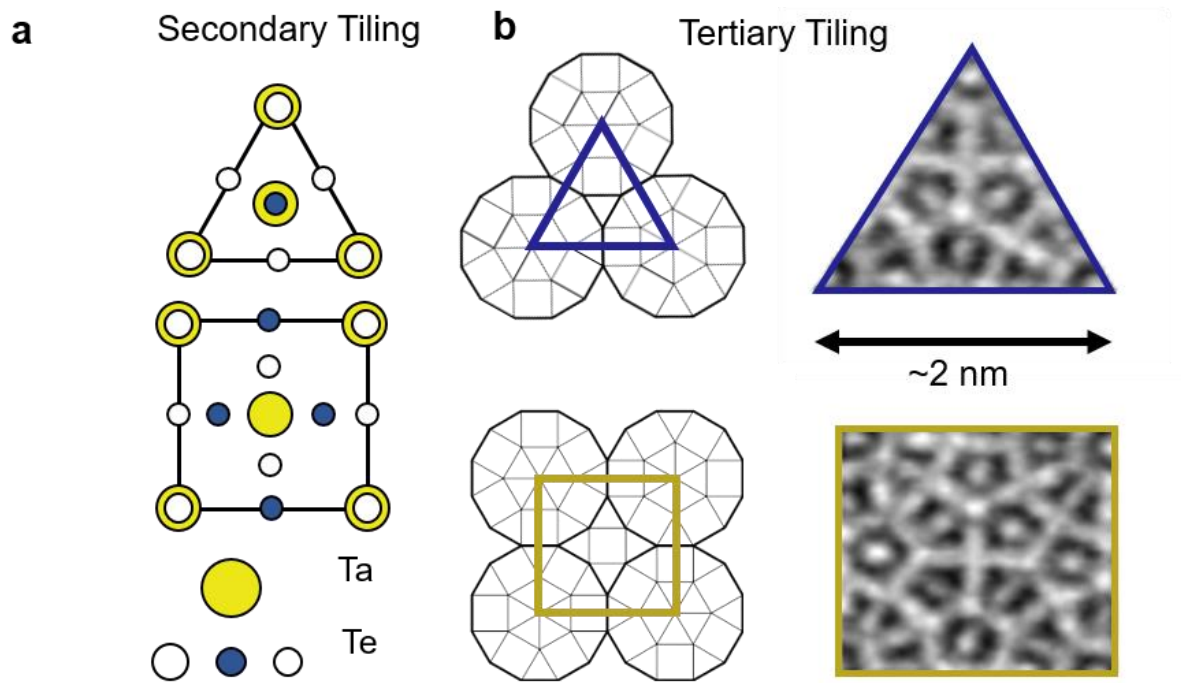
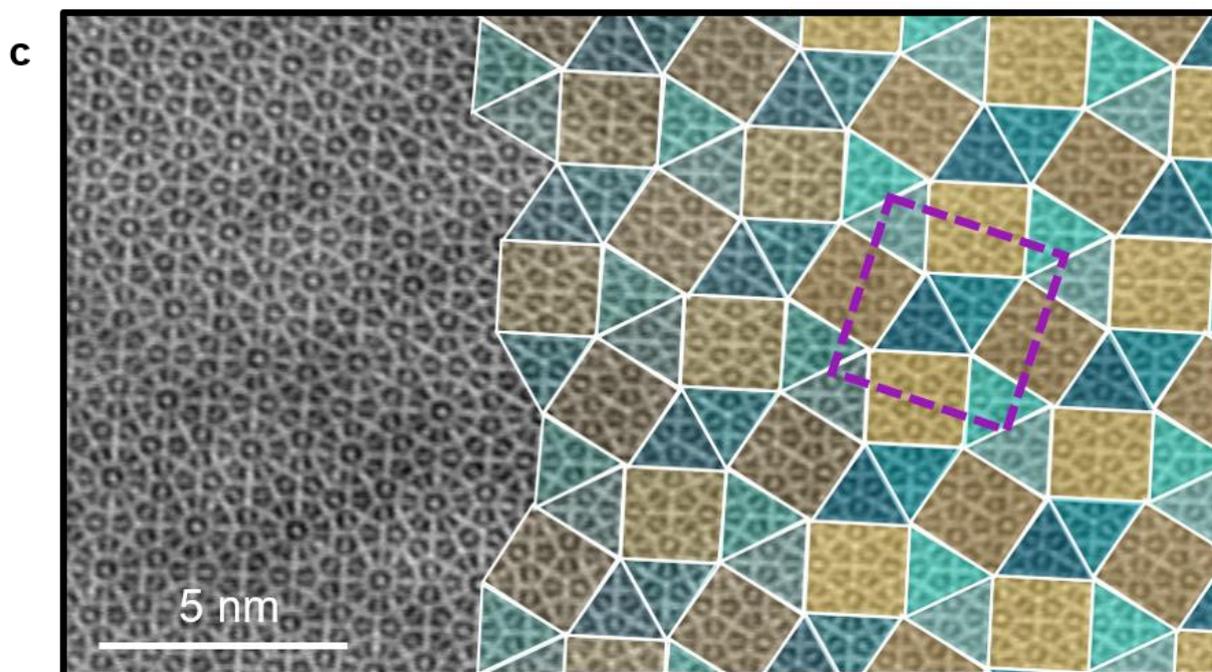

**Fig 3. Tiling of Ta$_{181}$Te$_{112}$ a**, Secondary tiling with atoms and **b,** tertiary triangle and square tiles used to tile the Ta$_{181}$Te$_{112}$ lattice. **c**, tiled image of Ta$_{181}$Te$_{112}$ (unit cell highlighted in purple).

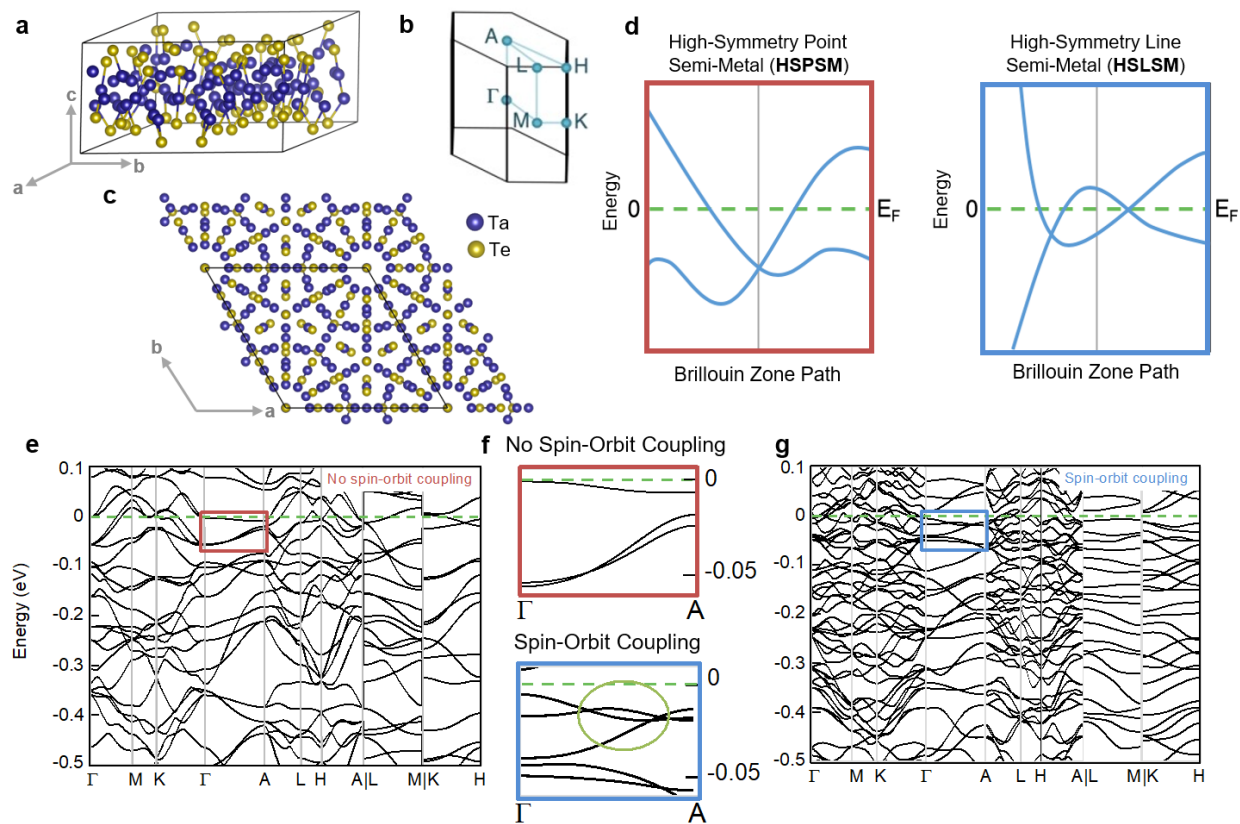

**Fig 4. Electronic Structure of the QC Approximant. a**, In-plane view **b**, Brillouin zone and **c**, Plane-view of the crystalline approximant $Ta_{21}Te_{13}$. **d,** Schematic of high-symmetry point semi-metal (HSPSM) and high-symmetry line semi-metal (HSLSM) band structures. **e**, Calculated band structure without spin-orbit coupling and **g,** with spin-orbit coupling of bulk $Ta_{21}Te_{13}$. **f**, Γ-A section of the band structure showing HSPSM and HSLSM within calculated structure.